# Crystallographic-dependent bilinear magnetoelectric resistance in a thin WTe$_2$ layer

Tian Liu,[*] Arunesh Roy, Jan Hidding, Homayoun Jafari, Dennis K. de Wal, Jagoda Slawi´nska, Marcos H.D. Guimar˜aes, and Bart J. van Wees Zernike Institute for Advanced Materials, Nijenborgh 4, 9747 AG Groningen, The Netherlands

(Dated: October 2, 2023)

# Abstract

The recently reported Bilinear Magnetoeletric Resistance (BMR) in novel materials with rich spin textures, such as bismuth selenide ($Bi_2Se_3$) and tungsten ditelluride ($WTe_2$), opens new pos-sibilities for probing the spin textures via magneto-transport measurements. By its nature, the BMR effect is directly linked to the crystal symmetry of the materials and its spin texture. There-fore, understanding the crystallographic dependency of the effect is crucial. Here we report the observation of crystallographic-dependent BMR in thin $WTe_2$ layers and explore how it is linked to its spin textures. The linear response measured in first harmonic signals and the BMR measured in second harmonic signals are both studied under a wide range of magnitudes and directions of magnetic field, applied current and at different temperatures. We discover a three-fold symmetry contribution of the BMR when current is applied along the a-axis of the $WTe_2$ thin layer at 10 K, which is absent for when current is applied along the b-axis.

I. INTRODUCTION

The study of spin dynamics in two-dimensional (2D) materials is promising for both fun-damental research and potential industrial applications [1–3]. In particular, the rich spin texture of $WTe_2$ has high potential for spin-active components in spintronic circuits, as indicated via spin- and angle-resolved photoemission spectroscopy (SR-ARPES) measure-ments [4–8]. Room temperature spin charge conversion was reported in $WTe_2$/graphene heterostructures and the direction of the generated non-equilibrium spin accumulation can be controlled via geometry designs [9–11]. However, the complex nature of the spin structure of $WTe_2$ still remains largely undercovered, while the material is predicted to be a type-II Weyl semimetal [12–15], hosting novel phenomena like extremely large magnetoresistance (XMR) [16], non-linear Hall effect [17, 18], spin momentum locking [19] and quantum spin Hall effect [20–22].

To unveil the potential of $WTe_2$ in spintronics, an essential step is to explore its spin texture and the associated spin dynamics under electric field and magnetic field, applied along different crystal axes. Spin polarized bands of $WTe_2$ has been measured by SR-ARPES and analyzed with DFT calculations, and spin polarized Fermi pockets of $WTe_2$ have been confirmed [6, 7]. As a step further, electrical probing of the the spin texture of materials is practical for measuring the direct response of the spin-texture-dependent effects under various experimental conditions and tailored to specific device geometries.

* tian.liu@rug.nl

For angular dependent magnetoresistance (ADMR) measurements in a bulk system, there are two relevant types of measurements which depend on the spin dynamics. The first type, i.e. spin accumulation related ADMR, is measured in the linear response. In the linear response, spin accumulation can be generated by a charge current, e.g. via spin Hall effect in metallic materials [23] or via Rashba–Edelstein effect in systems with k-dependent spin textures [24]. While the non-equilibrium spin accumulation is generated by a charge current, the reciprocal process is always present, where the spin current is converted to a charge voltage via the reciprocal effects, e.g. inverse spin Hall effect [25, 26] and inverse Rashba–Edelstein effect [27, 28]. When the spin accumulation is dephased by magnetic fields perpendicular to the direction of the spins, it results in Hanle magnetoresistance (HMR) [29]. Alternatively, the spin accumulation can also be controlled by the interaction with the magnetization of an adjacent layer, and it leads to spin Hall magnetoresistance (SMR) [30, 31].

The second type of the spin related ADMR is in the second order response, where the ADMR is linked to the spin current associated with the charge current instead of spin accumulation, as in second order a net spin current is allowed and it can be modulated by the applied magnetic fields. As a spin current is not allowed in the linear response in the spin orbit system by time reversal symmetry, this type of ADMR is not allowed in the first order. An example of the second type spin related ADMR is the bilinear magnetoelectric resistance (BMR) [32]. Therefore, BMR can be detected by applying a magnetic field, which creates unbalanced opposite spin directions as a result of Zeeman energy splitting, leading to a conversion from the spin current to a charge current.

BMR is a powerful tool to study spin textures of novel materials via electrical transport measurements. One advantage is that the spin texture can be directly probed via magneto-transport measurement in a material itself, without the need of fabricating a heterostructure to include other materials or magnetic contacts for spin injection/detection. It was reported that the BMR scales linearly with both electric field and magnetic field in the second order response [32]. Focusing on the second order, the resistance of the system can be described

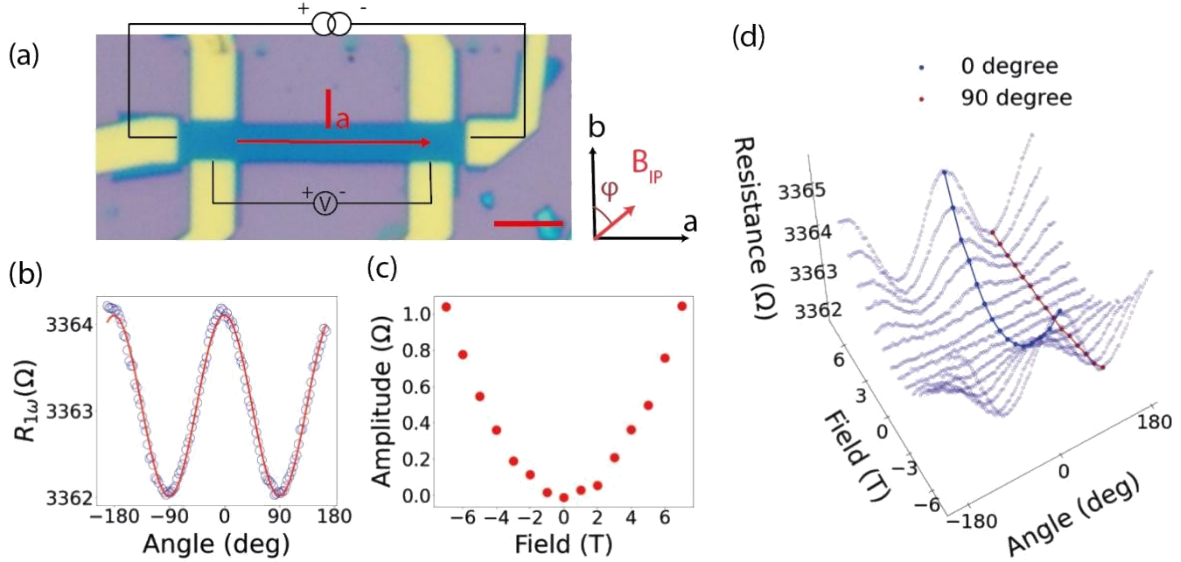

FIG. 1. Device geometry and measurement results for the first harmonic resistance measured at 300K. a). An optical microscope graph and the circuit for the flake resistance measurements along the long edge (a-axis); The bottom right scale bar in red is 5 μm long. b). Magnetoresistance in the linear response under in-plane magnetic field at the angle φ. The data is fitted with Equation (1); c). The fitted amplitude for magnetic field from -7T to 7T; d). The full dataset of MR under various magnetic fields. The modulation of the MR appears at 0° and 180°, when the field is perpendicular to the current. At 90° (-90°) when the field is parallel (anti-parallel) to the current, the MR does not scale quadratically with the field.

as:

$$R(I,B) = R_0 + \Delta R \cdot I \cdot B, \qquad (1)$$

where $R_0 = V/I$ is the resistance at zero magnetic field and low current and $\Delta R$ is the BMR coefficient. The BMR effect is sensitive to specific crystallographic orientation of the material [32]. It was reported that BMR measurements on $WTe_2$ are linked to Fermi surface topology and convexity [33]. However, the crystal-axis dependent spin texture of $WTe_2$ has not yet been revealed via the BMR effect. As the bc plane is the only mirror plane of the crystal and the ac plane is not a mirror plane [34], a crystallographic-dependent BMR is allowed and expected.

In this work, we fabricated three $WTe_2$ Hall bar devices (sample A, B and C) with different thickness and measured their electronic transport properties. We only present the measurement results of the sample A (9-nm WTe$_2$ flake on top of SiO$_2$) in the main text, as it shows the most pronounced ADMR signals. The optical images of the three samples and the measurement results can be found in the SI. The WTe$_2$ flakes were exfoliated from orthorhombic WTe$_2$ crystals (obtained from HQGraphene) on SiO$_2$ substrates. The exfoliation was performed in a nitrogen atmosphere with O$_2$ < 0.1 ppm and H$_2$O < 0.5 ppm. The contacts were designed and written following standard electron-beam lithography procedures. Before depositing Au/Ti contacts (75nm/5nm in thickness) on top of the WTe$_2$ via an electron-beam evaporation system, mild argon ion milling was performed for 20s to remove the possible degraded surface of WTe$_2$, as the flake was exposed to air for tens of seconds before loading to the metal deposition system. After the metal depostion, a lift-off procedure was followed and the sample was emerged in acetone solution before its loading into a different glove box filled with nitrogen (the oxygen level below 0.1%; H$_2$O level below 2.8mbar), where the sample was spin coated with PMMA. Through a standard E-beam lithography step, a Hall bar shape of PMMA layer was remained on top of the flake, acting as an etching mask. Then the WTe$_2$ flake was etched into the Hall bar shape by CF$_4$ reactive-ion etching, following the natural exfoliated direction of the samples. For the sample presented in the main text of this paper, the long edge of the flake is aligned to the a-axis of WTe$_2$, as indicated in Fig 1.a. The crystal orientation was determined by polarized Raman measurements directly on this device, as shown in Supporting Information (SI) [35]. After etching, the PMMA mask was removed and an extra PMMA layer (270 nm thick) was spin coated to protect the sample during sample bonding and electrical measurements.

II. MEASUREMENTS

In our electrical measurements, we applied AC currents along different crystal axes of WTe$_2$ and separate different harmonics via standard lock-in techniques. The voltage response is composed of different orders and are expanded as: V (t) = R$_1$I (t) + R$_2$I$^2$ (t) + · · · [36], where R$_i$ is the ith-order response [37] to the applied AC current I (t) [38]. We measure first harmonic signal (i.e. linear response R$_{1\omega}$) and second harmonic signal (R$_{2\omega}$) for both longitudinal voltage (Fig.1 - Fig.4) and transverse voltage, as shown in the supporting information (SI) [35]. We focused on longitudinal and transverse resistance measurements on a Hall bar device made from a 9-nm thick WTe$_2$ flake, as a function of the direction and mag-nitude of the electric and magnetic fields. We also perform similar measurements on other two samples which show overall similar results for the linear response. The measurements are shown in the SI [35].

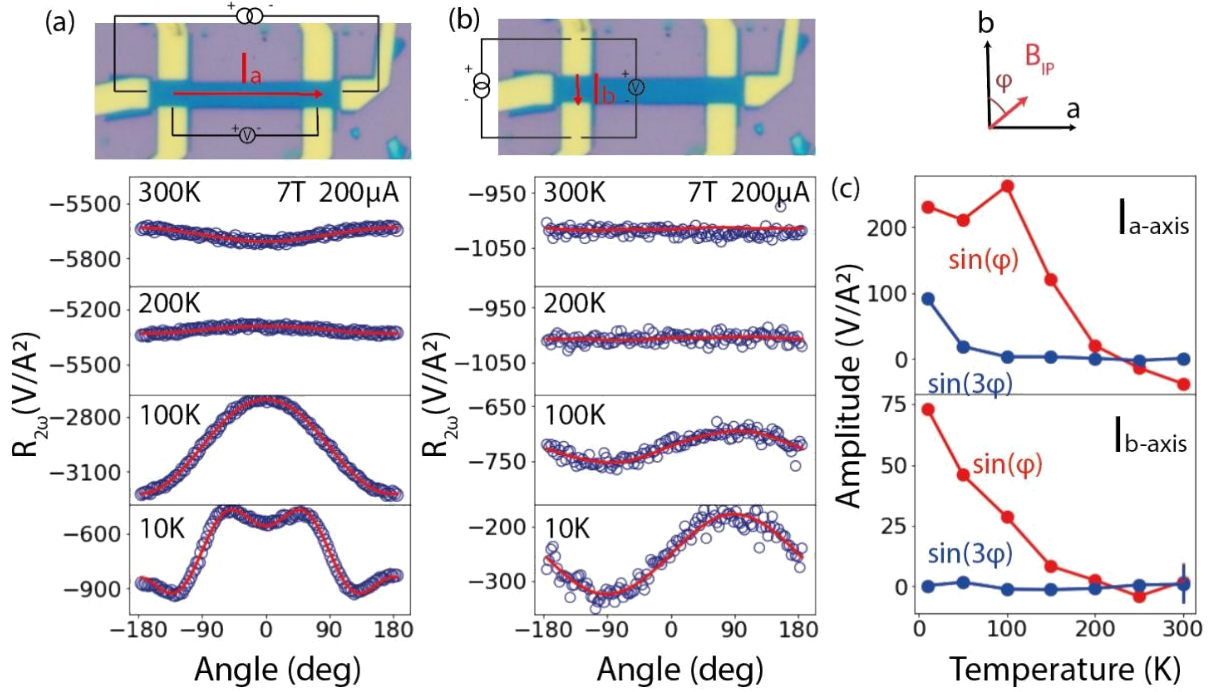

FIG. 2. The second harmonic longitudinal resistance measurements at different temperatures. a). Four-terminal longitudinal second harmonic resistance along the a-axis. b). Two-terminal longitudinal second harmonic resistance along the b-axis. Fitting amplitudes of the sin(ϕ) (red) and sin(3ϕ) (blue) for the longitudinal second harmonic resistance with the current along the a-axis (top panel (c)) and b-axis (bottom panel (c)). The error bars that are not visible, are smaller than the point size.

A typical four-terminal longitudinal voltage measurement is shown in Fig 1.a, in this configuration an AC current of 200 μA is applied along the a-axis of WTe$_2$. The data were measured at 300 K. When an in-plane magnetic field is applied at an angle φ with respect to I, we observe an angular dependent magnetoresistance (ADMR) which shows a periodicity of 180° in the linear response (Fig.1b). The measured ADMR is fitted with a sine function:

$$R_{1\omega} = A \sin(2\varphi + \pi) + B, \qquad (2)$$

where A is the amplitude of the ADMR, and B is the background resistance. We found the fitted amplitude A depends on the magnetic field quadratically, as shown in Fig.1c, which is consistent with the literature [33]. At 7 T, it shows an amplitude A = 1.044 ± 0.008 Ω, and the total modulation of the ADMR (2A) is 0.06% of the flake resistance at room temperature. We take a linear background into account during the fitting, which we believe it comes from the capacitive couplings in the circuit.

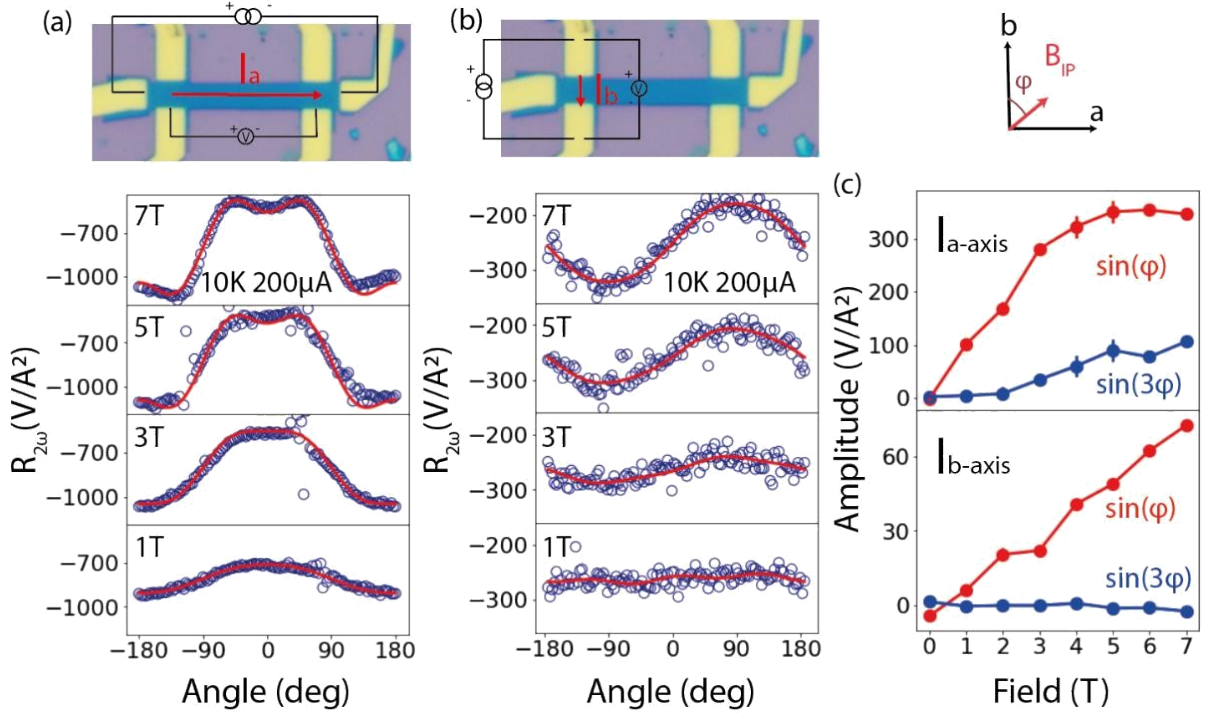

FIG. 3. The second harmonic longitudinal resistance measurements under different magnetic fields. a). Four-terminal longitudinal second harmonic resistance along the a-axis. b). Two-terminal longitudinal second harmonic resistance along the b-axis. Fitting amplitudes of the sin(ϕ) (red) and sin(3ϕ) (blue) for the longitudinal second harmonic resistance with the current along the a-axis (top panel (c)) and b-axis (bottom panel (c)). When the error bars are not visible, they are smaller than the point size.

Interestingly, by ADMR measurements under a range of magnetic fields in different magnitude from -7 T to 7 T (as shown in Fig.1d), we find that the first order ADMR is strongly sensitive to the angle between the applied current direction and the magnetic field direction. The ADMR reaches the maximum value and scales quadratically with magnetic field, when the field is perpendicular to the current (as shown by the blue dots data measured at φ = 0° in Fig.1d). There is no quadratic dependence when the field is along the current (as shown by the red dots data measured at φ = 90° in Fig.1d), while a almost linear change of the resistance on the field is observed, although it is unclear where the background signal on the field comes from. The ADMR shown in the first harmonic signal $R_{1\omega}$ is independent on the crystal axes. These results indicate the ADMR in this linear response comes from a mechanism which is independent of crystallographic orientations and it is different from the following discussion of BMR.

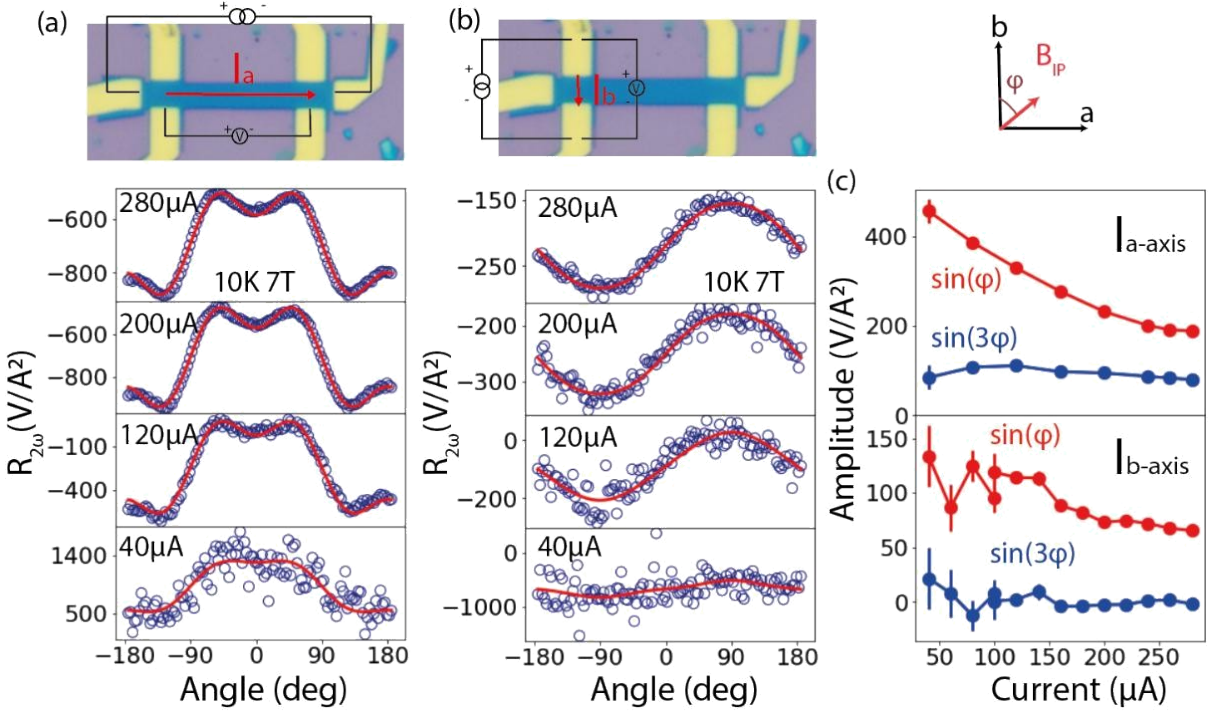

FIG. 4. The second harmonic longitudinal resistance measurements with different currents. a). Four-terminal longitudinal second harmonic resistance along the a-axis. b). Two-terminal longitudinal second harmonic resistance along the b-axis. Fitting amplitudes of the sin(ϕ) (red) and sin(3ϕ) (blue) for the longitudinal second harmonic resistance with the current along the a-axis (top panel (c)) and b-axis (bottom panel (c)). When the error bars are not visible, they are smaller than the point size.

A possible explanation to this ADMR at room temperature is the Hanle magnetoresis-tance (HMR). For HMR, the spin direction, and corresponding direction component of the spin accumulation which is perpendicular to the magnetic field is dephased. The effect scales quadratically with the magnetic field. The HMR type of effect generally leads to an increase of the resistance which fits into our observation in Fig.1b. However, in special cases e.g. in the chiral system the HMR type of effect could also lead to a decrease of the resistance [39]. This quadratic dependence matchs with our observation as shown in Fig.1c, implying the spin direction of the spin accumulation is along the current. As a possible explanation of the HMR, we resort to the symmetry analysis. From the symmetry point of view, the 9-nm thick flake can be associated with space groups 6, 11 or 31 and the applied current is either along the a-axis or b-axis of the crystal. The unconventional spin Hall effects where the applied current direction, electric field and spin polarizations are mutually collinear are not allowed by these space groups [40]. Therefore such spin Hall effects are ruled out from the possible origins.

In the SI [35], we also show a reciprocity check for the transverse resistance: after interchanging the current source and the voltage probe, the measurement results are the same. In equation (2), a phase 'π' is introduced for making the fitted amplitude positive for the 300 K ADMR. By adding this phase we define this as a positive magnetoresistance, due to the resistance increase while the field is perpendicular to the current.

We now discuss the second harmonic measurements. The BMR effect is measured via the second harmonic signals $R_{2\omega}$, and it shows strong crystallographic dependence at temperatures below 50K (deviations from a pure (co) sinusoidal dependence). Fig.2a and Fig.2b show measurements for the current applied along a-axis and b-axis respectively. Clear differences on the BMR amplitude and periodicity are observed. The transverse $R_{2\omega}$ is also measured and shown in SI [35]. For the longitudinal $R_{2\omega}$, in contrast to previous reports [33] where only a $\sin(\varphi)$ component was observed, here we measured an additional $\sin(3\varphi)$ component when the current is applied along the a-axis of the $WTe_2$ crystal. Our data is fitted by:

$$R_{2\omega} = A_1\sin(\varphi + \pi) + A_2\sin(3\varphi + \pi/2) + B \qquad (3)$$

$A_1$ is the amplitude for the $\sin(\varphi)$ part and $A_2$ is the amplitude for the $\sin(3\varphi)$ contribution. A φ-dependent background B is always present and it depends on the temperature, field and the applied current (see SI [35] for details). A similar equation is used to fit the data which was measured when the current was applied along the b-axis:

$$R_{2\omega} = A_1\sin(\varphi + \pi/2) + A_2\sin(3\varphi) + B \qquad (4)$$

Note here a different phase π/2 is used due to a π/2 change on the applied current direction. The biggest amplitude $A_2$ is at 10 K (the lowest measured temperature) when a current is applied along the a-axis, while the $A_1$ saturates for temperatures lower than 100 K. We notice that the sin(3ϕ) modulation is similar to the DC transverse resistance measurement from Li et al. [41], where the harmonic separation is lacking, and it is not clear if the sin3phi component is in the linear or non-linear response. We report this sin(3ϕ) contribution in the longitudinal resistance measurement and identify it in the second harmonic for the first time.

We also measured the longitudinal $R_{2\omega}$ for different magnitudes of magnetic field at 10 K for a fixed applied current of 200 µA, as shown in Fig.3a for the current applied along a-axis and in Fig.3b for the current applied along b-axis. These data are fitted with Eq.(3) and Eq.(4) respectively and the fitting amplitudes are plotted in Fig.3c and Fig.3d. In the BMR studies of both $Bi_2Se_3$ [32] and $WTe_2$ [33], the effect scales linearly with the field with a periodic angular dependency of 2π. We observe the similar results, i.e. the linear dependency of the longitudinal BMR on the field, when the current is applied along the a-axis, as shown in Fig.3d. In this case, no significant sin(3φ) contribution is observed. This linear dependence of BMR on the field is not observed anymore when the current is applied along the b-axis, especially for the sin(φ) component. At 10 K and with an applied current of 200µA, both the sin(φ) component and the sin(3φ) component of BMR saturate for the fields higher than 5T.

We also show the current dependence of BMR in Fig.4a for the current applied along the a-axis and in Fig.4b for the current applied along the b-axis. These data are fitted with Eq. (3) and Eq.(4) respectively and the results are shown in Fig.4c and Fig.4d. The BMR resistance $R_{2\omega}$ is supposed to be constant with the current when it is expressed in the unit of $V/A^2$ (and it is supposed to be linear with the current when it is in the unit Ω). However, we observed a higher BMR signal at low currents, as shown in Fig.4c and Fig.4d. In the SI [35], we plot measurements with different currents at 10 K without a magnetic field. We found a resistance change of 5% from 20 µA to 200 µA, and therefore the 50% drop of the amplitude of $R_{2\omega}$ in a-axis cannot be explained solely by heating. The amplitude of $R_{2\omega}$ in Fig.2 does not highly depend on the temperature when below 100K as well.

III. DISCUSSION

To give a tentative understanding of our results, in particular the different periodicities of the first harmonic (linear) response and the second order BMR, we note that the role of the magnetic field is different. Similar to the conventional Rashba Edelstein effect, in linear response the charge currents induce shifted distribution which produces a spin accumulation (note that there is no spin current generated in the linear regime). Similar to the inverse REE effect, the spin accumulation is converted back in to a current, which can increase/decrease the resistance. The applied magnetic field, when it is perpendicular to the direction of the spin accumulation, will induce spin precession, and thus reduce the spin accumulation. Thus the resistance will be modified accordingly. Note that the dephasing will be the same when the magnetic field is reversed by 180 degrees, and therefore the resistance modulation in the linear regime will have a 180 degrees' periodicity of the magnetic field direction, and the amplitude will depend on $B^2$.

The origin of the first harmonic ADMR remains an open question, on whether it arises from HMR. Future experiments to verify this could use spin pumping measurements in a YIG (yittrum iron garnet)/WTe$_2$ structure, where a pure spin current can be addressed in a more well-defined way, and the spin to charge conversion can be separated from the charge to spin conversion. This would help identifying the mechanism of the reported first harmonic ADMR.

The origin of the second order BMR is different. Here we can give a tentative explanation by noting that in the non-linear regime a spin current can be generated. Without a magnetic field there is no symmetry breaking, and no charge current will evolve. However, applying a magnetic field parallel to the spin polarization, an asymmetry can be induced by the Zeeman energy between spins propagating in the positive k direction and those in the negative k direction. This will result in a charge current in the second order, whose strength and direction will depend linearly on the magnetic field. However, the details of this mechanism will strongly depending on the specific spin texture. We have calculated the spin texture (as shown in the SI [35]), where the predominate periodicity is 360 degrees. Corresponding to this the periodicity, when rotating the magentic field direction, the BMR periodicity will be 360 degrees. However, we did not find an explanation for the 3-fold symmetry of the BMR signal at the low temperature. This remains to be explored for future research.

The exact calculation of BMR becomes very challenging for mutilayer WTe$_2$ due to the complication of a finite temperature and numerous spin subbands. Besides the non-zero temperature and the sophisticated spin subbands, we would like to also point out that the spin texture itself could already change its orientation in high magnetic fields (even at around 10 T) and the Zeeman energy could play a role on the measured second order resistance [42]. Normally a Zeeman energy will not affect the transport of a 2D system, as the density of states will decrease canceling the lifted speed of charge carriers. However, in a quasi-2D system, the role of Zeeman energy becomes non-trivial. To fully understand the origin of the $\sin(3\phi)$ BMR contribution, a dedicated investigation into the microscopic details of thin $WTe_2$ is needed for future investigations.

In summary, we report interesting transport measurements on $WTe_2$ devices with a Hall bar geometry. The first harmonic ADMR is sensitive to the magnetic field perpendicular to the applied current and it is crystal axis independent. The origin of this ADMR is to be explored. We also observe a second harmonic ADMR originating from the BMR, which is strongly crystallographic-dependent. A $\sin(3\phi)$ contribution of BMR is reported for the first time and we believe it is linked to the spin texture of the 9-nm $WTe_2$ thin flake. Those observations are relevant for understanding the spin texture and the nature of spin current dynamics in thin $WTe_2$ layers and pave the way for further spintronic applications utilizing the rich spin texture in $WTe_2$ and other relevant 2D materials.

ACKNOWLEDGMENTS

The authors thank Prof. Saroj Dash and Dr. Bing Zhao for helpful discussion on the device fabrication. The authors thank prof. Maria Loi and Eelco Tekelenburg for granting us access to their polarized Raman setup and necessary helps. The authors would also like to thank Dr. Alexey Kaverzin for the discussion on the resistance calculation, Dr. Julian Peiro and Dr. Si Chen for interesting discussions which spark the initial idea of this project. This project has received funding from the Dutch Research Council on Matter (FOM, now known as NWO-I) as a part of the Netherlands Organization for Scientific Research (NWO), the European Unions Horizon 2020 research and innovation programme under grant agreement No 696656 and 785219 (Graphene Flagship Core 1, Core 2 and Core 3) and Zernike Institute for Advanced Materials. MHDG acknowledges support from NWO STU.019.014. J.S. acknowledges the Rosalind Franklin Fellowship from the University of

Groningen.

---